\documentclass[aps,preprint,showpacs,amsmath,amssymb,prd,nofootinbib]{revtex4}
\usepackage{amssymb,graphics,graphicx}

\newcommand{\be}{\begin{equation}} \newcommand{\ee}{\end{equation}}
\newcommand{\bea}{\begin{eqnarray}}
\newcommand{\eea}{\end{eqnarray}}
\begin{document}
\title
{\large \bf  Heterotic Discrete Flavor Model}
\author{Paul H. Frampton$^{1}$\footnote{paul.h.frampton@gmail.com},
Chiu Man Ho$^{2,\,3}$\footnote{chiuman.ho@vanderbilt.edu}
and Thomas W. Kephart$^{2}$\footnote{tom.kephart@gmail.com}}

\affiliation{
$^{1}$Department of Physics and Astronomy, University of North Carolina,
Chapel Hill, NC 27599, USA\\
$^{2}$Department of Physics and Astronomy, Vanderbilt University,
Nashville, TN 37235, USA \\
$^{3}$Department of Physics and Astronomy, Michigan State University, East Lansing, MI 48824, USA}


\date{\today}

\begin{abstract}
We present an extended 331 model with $T'$ discrete flavor symmetry that
simultaneously explains the need to have exactly three generations and
provides acceptable quark and lepton masses and mixings. New fermionic
states and gauge bosons are predicted within the reach of the LHC.
We discuss the relevance to the 126 GeV scalar discovered at the LHC.
\end{abstract}
\maketitle

\section{Introduction}

The particle at 126 GeV has now been established at the LHC to be
consistent with
the standard model (SM) Higgs~\cite{ATLAS,CMS}, with the correct
coupling strengths
to fermions \cite{coupling}, CP property \cite{CP},
and spin \cite{spin}.
What remains is a definitive extension of the standard model to
eliminate our lack of
understanding of fermion masses and mixing angles.
In addition, while chiral anomaly
cancellation restricts the combination of complex irreducible
representations (irreps)
that fermions can inhabit, but allow SM generations plus
right-handed singlet neutrinos,
the number of generations is not constrained by anomaly cancellation.
Unitarity of the Pontecorvo-Maki-Nakagawa-Sakata (PMNS)
~\cite{Pontecorvo:1957cp} and
Cabibbo-Kobayashi-Maskawa (CKM)~\cite{Cabibbo:1963yz}
matrices would tell us that
the first three generations can be sufficient unto
themselves (or at least decoupled from
additional fermions if they exist), while lack of unitarity
would imply more fermions,
either in extensions of the first three generations or in additional generations.\\

A number of models with discrete symmetries have been proposed to explain the
fermion masses and mixings. For recent reviews,
see~\cite{Altarelli:2010gt}
and~\cite{Ishimori:2010au}.  One model where many
of the masses and mixings
are calculable is the $T'$ model~\cite{Frampton:1994rk}
which has been explored
in detail in
~\cite{Aranda:2000tm,Frampton:2007et,Frampton:2008bz,Eby:2008uc,Eby:2009ii,Ho:2010yp,Frampton:2010uw,Eby:2011ph,EbyThesis},
where $T'$ is the binary tetrahedral group which is economical
in the sense that it has only 24 elements
yet still has sufficient irreps
$(\mathbf{1_1}, \,\mathbf{1_2},\, \mathbf{1_3},\, \mathbf{2_1},\,\mathbf{2_2},\,\mathbf{2_3},$ and $\mathbf{3})$ to
arrange masses and mixings in agreement with the experimental
data. The top quark is
naturally split off from the light quarks by the choice of
embedding in $T'$, the
Cabibbo angle is directly calculable, etc. We tabulate
the Kronecker products for the irreducible
representations of $T'$ \cite{Frampton:2007et} in Table I.

\begin{table}
\centering
\begin{tabular}{||c||c|c|c|c|c|c|c||}
\hline\hline
 & $~1_1~$  & $~1_2~$ & $~1_3~$ & $2_1$ & $2_2$ & $2_3$ & $~3~$ \\
\hline\hline
$~1_1~$ & $1_1$ & $1_2$ & $1_3$ & $2_1$ & $2_2$ & $2_3$ & $3$  \\
\hline
$~1_2~$ & $1_2$ & $1_3$ & $1_1$ & $2_2$ & $2_3$ & $2_1$ & $3$  \\
\hline
$~1_3~$ & $1_3$ & $1_1$ & $1_2$ & $2_3$ & $2_1$ & $2_2$ & $3$  \\
\hline
$~2_1~$ & $2_1$ & $2_2$ & $2_3$ & $1_1 + 3$ & $1_2 + 3$ & $1_3 + 3$ & $2_1 + 2_2 + 2_3$ \\
\hline
$~2_2~$ & $2_2$ & $2_3$ & $2_1$ & $1_2 + 3$ & $1_3 + 3$ & $1_1 + 3$ & $2_1 + 2_2 + 2_3$ \\
\hline
$~2_3~$ & $2_3$ & $2_1$ & $2_2$ & $1_3 + 3$ & $1_1 + 3$ & $1_2 + 3$ & $2_1 + 2_2 + 2_3$ \\
\hline
$~3~$ & $~3~$ & $~3~$ & $~3~$ & $~2_1 + 2_2 + 2_3~$ & $~2_1 + 2_2 + 2_3~$ & $~2_1 + 2_2 + 2_3~$ &
$~1_1 + 1_2 + 1_3 + 3 + 3~$ \\
\hline\hline
\end{tabular}
\caption{~Kronecker products for the irreducible representations of $T'$.}
\end{table}

The standard model is insufficient for explaining why we have
three generations of fermions. Asymptotic freedom does
restrict the number, but does not fix it. A simple extended
model that does require three generations is the 331 model
\cite{Frampton:1992wt,Pisano:1991ee}, where the third
generation is assigned to the
$SU(3)_C\times SU(3)_L\times U(1)_X$ gauge group differently
from the first two, and it takes all three generations to
cancel the chiral anomalies. (This idea can be extended
to a class of models with the gauge group
$SU(3)_C \times SU(N)_L \times U(1)_X$ where for all
$N \geq 3$, anomaly cancellation requires three
generations \cite{Frampton:2012zp}.)
This again, as in the $T'$ model, allows for a heavy top
quark, plus now there are additional quarks and leptons in
the extended generations in addition to the leptoquark
gauge bosons due to the extended gauge symmetry.  While
explaining why there are three generations, the 331 model
has limited predictability of the fermion masses and mixings.

We will argue that the extension of the standard model with
$T'$ discrete flavor symmetry to a $331\times T'$ model avoids
some of the short comings of both the $T'$ and the 331 models.\footnote{Ref. \cite{S3}, which has a similar motivation as the current paper, considers $331\times S_3$. While $S_3$ is slightly simpler than $T'$, it accounts only for quark masses and mixing. However, $T'$ accounts for both of quark and lepton masses and mixing.} Our $331\times T'$ model retains all the predictions of the
original $\textrm{SM}\times T'$ model for fermion masses and mixings,
while three generations are still dictated by anomaly
cancellation as in the 331 model.

In the original 331 model, we need three $SU(3)_L$ triplet
scalars and one sextet scalar in order to accommodate
all the spontaneous symmetry breaking. However, as we will
show, the Higgs sector of the $\textrm{SM}\times T'$ model together
with only two additional $SU(3)_L$ triplet scalars are
sufficient for the extension to the $331\times T'$ model.
Thus, only a minimal extension of the Higgs sector is
required when the $\textrm{SM}\times T'$ model is extended to
the $331\times T'$ model and the complicated sextet
scalar can be avoided. In general, it is non-trivial
that the resulting model is still consistent when two
distinct models are merged together; and even if it can
be made so, many nontrivial extensions of the spectrum
may be required. On the contrary, the extension from
$\textrm{SM}\times T'$ to $331\times T'$ retains all the merits
of each of 331 and $\textrm{SM}\times T'$, while at the same time,
maintaining  consistency and simplicity.
Besides, many more predictions of the new model than of
either the 331 model or the $\textrm{SM}\times T'$ model are within
reach of the LHC. In this sense, the new $331\times T'$ model
is more than the sum of its parts and can justifiably be called a
``heterotic" model.

Before describing our model, we note that SUSY and
fourth-generation models are becoming disfavored by
the data, which makes the exploration of alternative
extensions of the SM more attractive.
(We note that a SUSY  $SU(5)\times T'$ model has been
studied~\cite{Chen:2007afa} as has a SUSY extended
$331\times T'$ \cite{Sen:2007vx}.)

\section{Particle Assignments}

Let us begin with the particle assignments under the
$SU(3)_C\times SU(3)_L\times U(1)_X$ gauge group and
$T'\times Z_2$ discrete symmetry of the $331\times T'$
model. The quantum numbers are written in the following manner:
\bea
\left(\,SU(3)_C,\, SU(3)_L, U(1)_X,\;
\mathbf{T'},\, Z_2\,\right)\,.
\eea

We embed the left-handed SM leptons in $SU(3)_L$ anti-triplets:
\bea
\begin{array}{ccc}
\left. \begin{array}{c}
\left( \begin{array}{c} E_e^+ \\ \nu_e \\ e^- \end{array} \right)_{L}  \\
\left( \begin{array}{c} E_\mu^+ \\ \nu_{\mu} \\
\mu^- \end{array} \right)_{L} \\
\left( \begin{array}{c} E_\tau^+ \\ \nu_{\tau} \\
\tau^- \end{array} \right)_{L} \\
\end{array} ~\right\}
L_L ~~  (1,\,\bar{3},\,0, \;\mathbf{3},\, +1)
&
\begin{array}{c}
~~~~~~ E_{e R}^+ ~ (1,\,1,\,1,\;\mathbf{1_3},\, +1) \\
~~~~~~ E_{\mu R}^+ ~ (1,\,1,\,1,\;\mathbf{1_2},\, +1) \\
~~~~~~ \,\,\,E_{\tau R}^+ ~ (1,\,1,\,1,\;\mathbf{1_1},\, +1)\,,
  \end{array}
\end{array}
\eea
where there is an additional charged heavy lepton per triplet
whose right-handed partners are in singlets as are the
right-handed partners of the SM charged leptons. 
Three right-handed neutrinos are included for use in the see-saw
mechanism as needed for masses and mixings:
\bea
\begin{array}{c}
 e^-_{R} ~ (1,\,1,\,-1,\;\mathbf{1_3},\, -1) \\
 \mu^-_{R} ~ (1,\,1,\,-1,\;\mathbf{1_2},\, -1) \\
 \tau^-_{R}~ (1,\,1,\,-1,\;\mathbf{1_1},\, -1)
\end{array}
&
\begin{array}{c}
~~~~~~ N^{(1)}_{R} ~ (1,\,1,\,0,\;\mathbf{1_1},\, +1) \\
~~~~~~ N^{(2)}_R ~ (1,\,1,\,0,\;\mathbf{1_2},\, +1) \\
~~~~~~ \,\,\,N^{(3)}_{R} ~ (1,\,1,\,0,\;\mathbf{1_3},\, +1)\,.
\end{array}
\eea
The 331 charges and irreps are dictated by anomaly
cancellation (except for the right-handed neutrinos).
The discrete group assignments are similar to and extend
those of the $\textrm{SM}\times T'$ model.

The SM quarks are also assigned as in the $\textrm{SM}\times T'$
model with extended generation quarks included naturally
as we increase the $SU(2)_L$ gauge symmetry to $SU(3)_L$:
\bea
\begin{array}{cc}
\left( \begin{array}{c} b \\ t \\ T \end{array} \right)_{L}
~ {\cal Q}_L ~~~~~~~~~~~~~~~ (3,\,\bar{3},\,\frac23,\;
\mathbf{1_1},\, +1)   \\
\left. \begin{array}{c} \left( \begin{array}{c} c \\ s
\\ S
\end{array} \right)_{L}
\\
\left( \begin{array}{c} u \\ d \\ D
\end{array} \right)_{L}  \end{array} \right\}
Q_L ~~~~~~~~~~ (3,\,3,\,-\frac13,\;\mathbf{2_1},\, +1)\,.
\end{array}
\label{qL}
\eea
Using the convention
$\rm{Tr} (\lambda_a\,\lambda_b) = 2\, \delta_{ab}$,
two of the generators for $SU(3)_L$ are given by
$\lambda_3= \rm{diag}\,\left(1, -1, 0\right)$
and
$\lambda_8 = \frac{1}{\sqrt{3}}\, \rm{diag}\,\left(1, 1, -2\right)$.
After the symmetry
$SU(3)_L\times U(1)_X$ breaks down to
$SU(2)_L\times U(1)_Y$,\, the hypercharge $Y$ is
\cite{Frampton:1992wt}:
\bea
\label{Y}
Y= 2\,X + \sqrt{3}\;\lambda_{8} \,,
\eea
and the electric charge is given by
$Q= T_{3} + \frac{Y}{2}$ where $T_3 = \frac12\,\lambda_3$.
The right-handed SM quarks are again as in the
$\textrm{SM}\times T'$ model, but the new heavy right-handed
quarks are arranged with $T_R$ in a $T'$ singlet and
$S_R$ and $D_R$ in a $T'$ doublet:
\bea
\begin{array}{c}
 \,T_R   ~~~~~~~~~~~~~~~ (3,\,1,\,\frac53,\;\mathbf{1_1},\, +1) \\
~ \left. \begin{array}{c} S_{R} \\ D_{R} \end{array} \right\}
{\cal D}_R ~~~~~~~ (3,\,1,\,-\frac43,\;\mathbf{2_1},\, +1) \\
\,\,t_{R} ~~~~~~~~~~~~~~~\, (3,\,1,\,\frac23,\;\mathbf{1_1},\, +1)   \\
\,~~~b_{R} ~~~~~~~~~~~~~~~\,(3,\,1,\,-\frac13,\;\mathbf{1_2},\, -1)  \\
\left. \begin{array}{c} c_{R} \\ u_{R} \end{array} \right\}
{\cal C}_R ~~~~~~~~ (3,\,1,\,\frac23,\;\mathbf{2_3},\, -1)\\
~~~\left. \begin{array}{c} s_{R} \\ d_{R} \end{array} \right\}
{\cal S}_R ~~~~~~~~ (3,\,1,\,-\frac13,\;\mathbf{2_2},\, +1)\,.
\end{array}
\label{qR}
\eea

We only need to extend the Higgs sector of the
$\textrm{SM}\times T'$ model by two $SU(3)_L$ triplet scalars in
order to accommodate the additional spontaneous symmetry
breaking needed in the 331 sector. Hence, we choose the
Higgses
$H_{\mathbf{3}},\, H'_{\mathbf{3}},\, H_{\mathbf{1_1}},\,
H_{\mathbf{1_3}}$ as in \cite{Frampton:2008bz} together
with the two extra scalars
$\Phi_{\mathbf{3}}$ and $\Phi_{\mathbf{1_1}}$:
\bea
&& H_{\mathbf{3}} ~~ (1,\,\bar{3},\,0, \;
\mathbf{3},\, +1) \nonumber \\
&& H'_{\mathbf{3}} ~~ (1,\,\bar{3},\,1, \;\mathbf{3},\, -1) \nonumber \\
&& H_{\mathbf{1_1}} ~ (1,\,\bar{3},\,0, \;\mathbf{1_1},\, +1) \nonumber \\
&& H_{\mathbf{1_3}} ~ (1,\,\bar{3},\,1, \;\mathbf{1_3},\, -1) \nonumber \\
&& \Phi_{\mathbf{3}} ~~\, (1,\,3,\,1, \;\mathbf{3},\, +1) \nonumber \\
&& \Phi_{\mathbf{1_1}} ~\, (1,\,3,\,1, \;\mathbf{1_1},\, +1) \,.
\eea
These six scalars can acquire their respective vacuum expectation values (VEVs) through the Higgs potential shown in Appendix A.

At this point, we can write down the most general Yukawa
lagrangian for the lepton sector
\begin{eqnarray}
{\cal L}_Y^{\textrm{leptons}}
&=&
\frac{1}{2} \,M_1 \,\bar{N}_R^{(1)} \,N_R^{(1)} + M_{23}\,
\bar{N}_R^{(2)}\, N_R^{(3)} \nonumber \\
& & + \,\Bigg\{\,
Y_{1} \,\left(\, \bar{L}_L\, N_R^{(1)}\, H_3 \,\right) + Y_{2} \,
\left( \, \bar{L}_L \,N_R^{(2)}\,  H_3
\,\right) + Y_{3}\,
\left(\, \bar{L}_L\, N_R^{(3)} \,H_3\, \right)  \nonumber \\
&& ~~~~~+
Y_e \,\left(\, \bar{L}_L \, e_R \,H'_3 \,\right)
+ Y_\mu  \,\left(\, \bar{L}_L \,\mu_R  \,H'_3 \,\right) +
Y_\tau \,\left(\, \bar{L}_L \,\tau_R \,H'_3 \,\right)
\nonumber \\
&& ~~~~~+
Y'_e \,\left(\, \bar{L}_L \, E_{eR} \,\Phi^{\ast}_3 \,\right)
+ Y'_\mu  \,\left(\, \bar{L}_L \,E_{\mu R}  \,\Phi^{\ast}_3 \,\right) +
Y'_\tau \,\left( \,\bar{L}_L \,E_{\tau R} \,\Phi^{\ast}_3 \,\right)\,
\Bigg\}
+
\textrm{h.c.}\,.
\label{Ylepton}
\end{eqnarray}
and for the quark sector
\begin{eqnarray}
{\cal L}_Y^{\textrm{quarks}}
&=& Y_t \,\left(\, \{{\cal \bar{Q}}_L\}_{\mathbf{1_1}}\,
\{t_R\}_{\mathbf{1_1}} \,H_{\mathbf{1_1}}\,\right)
+ Y_b \,\left(\,\{{\cal \bar{Q}}_L\}_{\mathbf{1_1}}\,
\{b_R\}_{\mathbf{1_2}}\, H_{\mathbf{1_3}} \,\right) \nonumber \\
&&
+ \,Y_{{\cal C}} \,\left( \,\{ \bar{Q}_L \}_{\mathbf{2_1}} \,
\{ {\cal C}_R \}_{\mathbf{2_3}} \,H^{'\ast}_{\mathbf{3}}\,\right)
+ Y_{{\cal S}} \,\left(\, \{ \bar{Q}_L \}_{\mathbf{2_1}}\,
\{ {\cal S}_R \}_{\mathbf{2_2}} \, H^{\ast}_{\mathbf{3}}\,\right)
\nonumber \\
&& + \,Y_T \,\left(\, \{{\cal \bar{Q}}_L\}_{\mathbf{1_1}}\,
\{T_R\}_{\mathbf{1_1}} \,\Phi^{\ast}_{\mathbf{1_1}}\,\right)
+ Y_{{\cal D}} \,\left(\, \{ \bar{Q}_L \}_{\mathbf{2_1}}\,
\{ {\cal D}_R \}_{\mathbf{2_1}} \, \Phi_{\mathbf{1_1}}\,\right)
\nonumber \\
&&
+\, \textrm{h.c.}\,.
\label{Yquark}
\end{eqnarray}

In this $331\times T'$ model, the neutrino sector is
unchanged relative to the $\textrm{SM}\times T'$ model. So
the predictions for the masses and mixings of the neutrino
sector in the $331\times T'$ model are exactly  those
predicted by the $\textrm{SM}\times T'$ model. The SM quark and
charged-lepton Yukawa terms are unchanged as well.
The masses of the charged heavy leptons
$E_{e}^{\pm}, \,E_{\mu}^{\pm}, \, E_{\tau}^{\pm}$ are
determined by the product of $Y'$ and
$\langle \Phi_3 \rangle$, which can be chosen at the TeV scale.
Furthermore, the masses of the heavy quarks
$T, S, D$ inherited from the 331 model could also be produced
at the TeV scale through the appropriate choices of new
Yukawa couplings and $\langle \Phi_{\mathbf{1_1}} \rangle$.
For instance, if we assume that the new Yukawa couplings are of
order unity, then we simply require
$\langle \Phi_{\mathbf{1_1}} \rangle$ to be TeV scale.

\section{Anomalies}

There are six types of triangle anomalies. Namely,
$SU(3)_C^3$,\, $SU(3)_C^2\,X$,\, $SU(3)_L^3$,\,
$SU(3)_L^2\,X$,\, $X^3$\, and \,$X$,\,
where the last anomaly is from a mixed gauge-gravity
triangle diagram. In the $331\times T'$ model, the left-handed
leptons are assigned in $SU(3)_L$ anti-triplets with the
same $SU(3)_C\times SU(3)_L\times U(1)_X$ quantum numbers
as those in the original 331 model. The first two and the
third generations of quarks also carry exactly the same
quantum numbers as those in the original 331 model. The
right-handed neutrinos are irrelevant for any anomaly
cancellations as they are singlets of
\,$SU(3)_C\times SU(3)_L\times U(1)_X$. Since the
right-handed charged leptons are singlets under
\, $SU(3)_C\times SU(3)_L$,\, they do not participate in the
\,$SU(3)_C^3$,\, $SU(3)_C^2\,X$,\, $SU(3)_L^3$,
\, $SU(3)_L^2\,X$\, anomaly cancellations.
Thus, cancellations of these anomalies proceed in the same way
as in the original 331 model, which require three generations.
For the \,$X^3$\, and\, $X$ anomalies, the contributions
from \,$e_R,\, \mu_R,\, \tau_R$ \, cancel those from
\,$E_{e R},\, E_{\mu R},\, E_{\tau R}$\, respectively
and this happens generation by generation. However, for
the rest of the \,$X^3$\, and\, $X$ \,anomalies,
cancellations proceed in the same way as in the original
331 model, which again require three generations.

Therefore, our new $331\times T'$ model retains all the
acceptable predictions for fermion masses and mixings
contained in the original $321\times T'$ model, while
three generations are still required by anomaly
cancellation as in the original 331 model.

It is possible that the discrete symmetry $T'$
originates from a discrete gauge
symmetry that is spontaneously broken.
The advantage of having a gauge origin
is that $T'$ could then be protected against
violations by quantum gravity
effects \cite{KraussWilczek}. But the introduction
of a new discrete gauge symmetry
may also imply the possibility of discrete
gauge anomalies \cite{IbanezRoss,LuhnRamond}.
The requirement of discrete gauge anomaly cancellations
leads to the discrete anomaly
conditions after the discrete gauge symmetry is
broken. It is remarkable that our model with $T^{'}$
symmetry is discrete anomaly free (see {it e.g.}
the second row of Table $2$ in \cite{Luhn}).

\section{Yukawa couplings to 126 GeV scalar}

The discovery of a resonance at $\sim 126$ GeV at the LHC
\cite{ATLAS,CMS}, strongly suggestive of the Higgs boson,
has naturally caused intense interest. Its preliminary properties
are consistent within errors with the Higgs particle
predicted by the minimal standard model.
The two-body decays which can be measured accurately
in the near future include $H\rightarrow\gamma\gamma$,
$H\rightarrow\bar{b}b$, and $H\rightarrow\bar{\tau}\tau$.
These are respectively sensitive to the Yukawa couplings
$Y_{H\bar{t}t}$ (through the top triangle contribution
which competes with the $W$-loop), $Y_{H\bar{b}b}$,
and $Y_{H\bar{\tau}\tau}$.

In the minimal standard model, the Yukawa couplings
$Y_{H\bar{f}f}$ appear in the simple form
$Y_{H\bar{f}f} \bar{f} f H$
so that they are proportional to the masses
\begin{equation}
Y_{H\bar{f}f} \propto m_f\,,
\label{proportional}
\end{equation}
where the proportionality constant is $1/v$ with $v$ being the vacuum expectation value of the Higgs field.
In the current renormalizable model with a non-trivial flavor symmetry
$G_F = T^{'} \times Z_2$, there are several Higgs
and the Yukawa couplings of the lightest one will generally deviate from the simple proportionality
of Eq. (\ref{proportional}). Such deviations may likely be small
but crucial to understanding how the group $G_F$ operates. One
may even say that if the conventional prediction of
Eq. (\ref{proportional}) would hold exactly at high precision,
then the renormalizable $G_F$ models would be disfavored.

The statements above are true for general renormalizable $G_F$ models.
To illustrate them, we focus on the present choice
$G_F = T^{\prime} \times Z_2$
\cite{Frampton:1994rk} and the minimal model as previously discussed in \cite{Frampton:2008bz}.
We shall concentrate only on the third-generation couplings
$Y_{H\bar{f}f}$ for $f=t,b,\tau$. Imposing strict renormalizability on the lepton lagrangian
allows nontrivial terms for the $\tau$ mass:
\begin{equation}
Y_\tau \,\left(\, \bar{L}_L \,\tau_R \,H'_3 \,\right)+
\textrm{h.c.}\,,
\label{Taulepton}
\end{equation}
where $H'_{\mathbf{3}}$ transforms as $H'_{\mathbf{3}}(1,\,\bar{3},\,1, \;\mathbf{3},\, -1)$.
The Yukawa couplings to the third-generation of quarks are contained in
\begin{eqnarray}
{\cal L}_{\textrm{3rd}}^{\textrm{quarks}}
&=& Y_t \,\left(\, \{{\cal \bar{Q}}_L\}_{\mathbf{1_1}}\,
\{t_R\}_{\mathbf{1_1}} \,H_{\mathbf{1_1}}\,\right)
+ Y_b \,\left(\,\{{\cal \bar{Q}}_L\}_{\mathbf{1_1}}\,
\{b_R\}_{\mathbf{1_2}}\, H_{\mathbf{1_3}} \,\right)
+ \textrm{h.c.}\,.
\end{eqnarray}

No $T^{'}$ doublet
($\mathbf{2_1}, \mathbf{2_2}, \mathbf{2_3}$) scalars have been added. This
allows a non-zero value only for $\Theta_{12}$. The other CKM
angles vanish, making the third generation stable
and $m_b > m_{s,d}$ as outlined in \cite{Frampton:1994rk}.
Such a model leads to the formula \cite{Frampton:2008bz} for the Cabibbo angle
\begin{equation}
\tan 2\Theta_{12} = \left( \frac{\sqrt{2}}{3} \right)\,,
\label{Cabibbo}
\end{equation}
or equivalently, $\sin \Theta_{12} = 0.218..$, which is
close to the experimental value $\sin \Theta_{12} \simeq 0.227$.
It can also lead to the successful relationship between neutrino mixing angles $\theta_{ij}$
\begin{equation}
\theta_{13} = (\sqrt{2})^{-1} \left| \frac{\pi}{4} - \theta_{23} \right|\,,
\label{neutrinomixing}
\end{equation}
which is in excellent agreement with the latest experiments \cite{EF}.

In such a model, the lightest Higgs $H$ is a linear combination of other Higgs:
\begin{equation}
H = a \,H_{\mathbf{1_1}} + b\, H_{\mathbf{1_3}} + c\, H'_{\mathbf{3}} + ...\,,
\label{lighthiggs}
\end{equation}
and the consequent Yukawa couplings are
\begin{equation}
Y_{H\bar{t}t} = a^{-1}\, Y_t, ~~~~ Y_{H\bar{b}b} = b^{-1}\, Y_b, ~~~~ Y_{H\bar{\tau}\tau} = c^{-1} \,Y_{\tau}\,.
\label{modeified}
\end{equation}
The VEV $v$ is shared between the $<H_{\alpha}>$ ($\alpha = \mathbf{1_1},\, \mathbf{1_3},\, \mathbf{3},...)$
irreps of $T^{\prime}$ and there
is no reason to expect $a=b=c=\dots $ so that
the proportionality of Eq. (\ref{proportional}) will generally be lost.
In fact, if Eq. (\ref{proportional}) remained exact, the only solution would
be a trivial one where all states transform as $\mathbf{1_1}$ of $T^{\prime}$
and the $G_F$ is inapplicable. The successes in \cite{Frampton:2008bz} and \cite{EF}
would, in such a case, be accidental.
On the other hand, if Eq. (\ref{proportional}) is inexact, the evaluations
of the coefficients $a,b,c,...$ can then be used to understand more
perspicuously the derivations of mixing angles for quarks and leptons
given respectively in \cite{Frampton:2008bz} and \cite{EF}, in a first
clear departure from the minimal standard model.

\section{Discussion}

One of the perpetual difficulties one encounters in constructing
models of fermionic flavor is the necessity of an extended Higgs
sector. Typically, if more symmetry is added, then more scalars are needed to
break that symmetry and generate the wanted structure of masses and
mixings. The model at hand is no exception, although it is somewhat more
attractive than average since all the scalars are in fundamental
irreps of the $SU(3)_L$ in the electroweak gauge group. (This is an
improvement on previous 331 models that required a {\bf 6} scalar of $SU(3)_L$.)
This simplification leads to fewer phenomenological problem, e.g., the $\rho$ parameter is
unchanged ent heterotic model provides new candidates for particles to
be discovered in the TeV range by, for example, the LHC.
As new gauge bosons, there are the bileptons familiar from the 331 model
which come both doubly-charged and singly-charged in the $SU(2)$
doublets $(Y^{++}, Y^{+})$ and
$(Y^{--}, Y^{-})$ with striking signatures in like-sign lepton
pairs; also there is a $Z^{'}$.
From muonium-antimuonium conversion experiments
\cite{Willmann}, a lower bound on the bilepton mass
$M_Y > 850$ GeV has been deduced, although this assumed
flavor diagonality of the bilepton couplings. More general
analysis of bilepton production at the LHC, with weaker assumptions,
appears in \cite{Meirose} and \cite{Borges}.

There are additional fermions beyond the standard model. These
include the exotic quarks $D$ and $S$ with charge $Q = -4/3$
and the $T$ with $Q = +5/3$. There are also three new charged
leptons, one per family, $E_{e}$, $E_{\mu}$, and $E_{\tau}$.
All of these additional states are predicted to be in the TeV or multi-TeV
range, accessible to the LHC especially at
its full energy and luminosity.

\begin{acknowledgments}

\noindent
The work of P.H.F. was supported in part by U.S.
Department of Energy Grant No. DE-FG02-05ER41418.
C.M.H. and T.W.K. were supported in part
by U.S. Department of Energy Grant No. DE-FG05-85ER40226.
\end{acknowledgments}

\appendix

\section{~~ Higgs Potential}

The VEVs for $H_{\mathbf{3}}$,\, $H'_{\mathbf{3}}$,\, $H_{\mathbf{1_1}}$,\, $H_{\mathbf{1_3}}$,\, $\Phi_{\mathbf{3}}$ and
$\Phi_{\mathbf{1_1}}$ can be obtained from minimizing the following Higgs potential:
\bea
V &=& \lambda_1\,\left(H^\dagger_{\mathbf{3}} H_{\mathbf{3}} - v^2_{\mathbf{3}} \right)^2
     +\, \lambda_2\,\left(H'^\dagger_{\mathbf{3}} H'_{\mathbf{3}} - v'^{\,2}_{\mathbf{3}}\right)^2
     +\, \lambda_3\,\left(H^\dagger_{\mathbf{1_1}} H_{\mathbf{1_1}} - v^{2}_{\mathbf{1_1}}\right)^2 \nonumber \\
&& + \,\lambda_4\,\left(H^\dagger_{\mathbf{1_3}} H_{\mathbf{1_3}} - v^{2}_{\mathbf{1_3}}\right)^2
     +\, \lambda_5\,\left(\Phi^\dagger_{\mathbf{3}} \Phi_{\mathbf{3}} - v''^{\,2}_{\mathbf{3}}\right)^2
     +\, \lambda_6\,\left(\Phi^\dagger_{\mathbf{1_1}} \Phi_{\mathbf{1_1}} - v'^{\,2}_{\mathbf{1_1}}\right)^2 \nonumber \\
&& +\, \lambda_7\,\left[ \left(H^\dagger_{\mathbf{3}} H_{\mathbf{3}} - v^2_{\mathbf{3}} \right) + \left(H'^\dagger_{\mathbf{3}} H'_{\mathbf{3}} -
v'^{\,2}_{\mathbf{3}}\right) \right]^2
+\, \lambda_8\,\left[ \left(H^\dagger_{\mathbf{3}} H_{\mathbf{3}} - v^2_{\mathbf{3}} \right) + \left(H^\dagger_{\mathbf{1_1}} H_{\mathbf{1_1}} -
v^{2}_{\mathbf{1_1}}\right) \right]^2 \nonumber \\
&& +\, \lambda_{9}\,\left[ \left(H^\dagger_{\mathbf{3}} H_{\mathbf{3}} - v^2_{\mathbf{3}} \right) + \left(H^\dagger_{\mathbf{1_3}} H_{\mathbf{1_3}} -
v^{2}_{\mathbf{1_3}}\right) \right]^2
+\, \lambda_{10}\,\left[ \left(H^\dagger_{\mathbf{3}} H_{\mathbf{3}} - v^2_{\mathbf{3}} \right)
+ \left(\Phi^\dagger_{\mathbf{3}} \Phi_{\mathbf{3}} -v''^{\,2}_{\mathbf{3}}\right) \right]^2 \nonumber \\
&& +\, \lambda_{11}\,\left[ \left(H^\dagger_{\mathbf{3}} H_{\mathbf{3}} - v^{\,2}_{\mathbf{3}} \right) + \left(\Phi^\dagger_{\mathbf{1_1}} \Phi_{\mathbf{1_1}} -v'^{\,2}_{\mathbf{1_1}}\right) \right]^2
+\lambda_{12}\,\left[ \left(H'^\dagger_{\mathbf{3}} H'_{\mathbf{3}} - v'^{\,2}_{\mathbf{3}} \right) + \left(H^\dagger_{\mathbf{1_1}} H_{\mathbf{1_1}} -v^{2}_{\mathbf{1_1}}\right) \right]^2 \nonumber  \\
&& +\, \lambda_{13}\,\left[ \left(H'^\dagger_{\mathbf{3}} H'_{\mathbf{3}} - v'^{\,2}_{\mathbf{3}} \right) + \left(H^\dagger_{\mathbf{1_3}} H_{\mathbf{1_3}} - v^{2}_{\mathbf{1_3}}\right) \right]^2
+\, \lambda_{14}\,\left[ \left(H'^\dagger_{\mathbf{3}} H'_{\mathbf{3}} - v'^{\,2}_{\mathbf{3}} \right) + \left(\Phi^\dagger_{\mathbf{3}} \Phi_{\mathbf{3}} -v''^{\,2}_{\mathbf{3}}\right) \right]^2 \nonumber \\
&&+\, \lambda_{15}\,\left[ \left(H'^\dagger_{\mathbf{3}} H'_{\mathbf{3}} - v'^{\,2}_{\mathbf{3}} \right) + \left(\Phi^\dagger_{\mathbf{1_1}} \Phi_{\mathbf{1_1}} -
v'^{\,2}_{\mathbf{1_1}}\right) \right]^2
+\, \lambda_{16}\,\left[ \left(H^\dagger_{\mathbf{1_1}} H_{\mathbf{1_1}} - v^2_{\mathbf{1_1}} \right) + \left(H^\dagger_{\mathbf{1_3}} H_{\mathbf{1_3}} - v^{2}_{\mathbf{1_3}}\right) \right]^2 \nonumber \\
&& +\, \lambda_{17}\,\left[ \left(H^\dagger_{\mathbf{1_1}} H_{\mathbf{1_1}} - v^2_{\mathbf{1_1}} \right)
+ \left(\Phi^\dagger_{\mathbf{3}} \Phi_{\mathbf{3}} -v''^{\,2}_{\mathbf{3}}\right) \right]^2
+\, \lambda_{18}\,\left[ \left(H^\dagger_{\mathbf{1_1}} H_{\mathbf{1_1}} - v^{2}_{\mathbf{1_1}} \right) + \left(\Phi^\dagger_{\mathbf{1_1}} \Phi_{\mathbf{1_1}} -v'^{\,2}_{\mathbf{1_1}}\right) \right]^2 \nonumber \\
&& +\, \lambda_{19}\,\left[ \left(H^\dagger_{\mathbf{1_3}} H_{\mathbf{1_3}} - v^2_{\mathbf{1_3}} \right)
+ \left(\Phi^\dagger_{\mathbf{3}} \Phi_{\mathbf{3}} -v''^{\,2}_{\mathbf{3}}\right) \right]^2
+\, \lambda_{20}\,\left[ \left(H^\dagger_{\mathbf{1_3}} H_{\mathbf{1_3}} - v^{2}_{\mathbf{1_3}} \right) + \left(\Phi^\dagger_{\mathbf{1_1}} \Phi_{\mathbf{1_1}} -v'^{\,2}_{\mathbf{1_1}}\right) \right]^2 \nonumber \\
&& +\, \lambda_{21}\,\left[ \left(\Phi^\dagger_{\mathbf{3}} \Phi_{\mathbf{3}} - v''^{\,2}_{\mathbf{3}} \right) + \left(\Phi^\dagger_{\mathbf{1_1}} \Phi_{\mathbf{1_1}} -v'^{\,2}_{\mathbf{1_1}}\right) \right]^2 \nonumber \\
&& +\, \alpha_{1} \, \left(H_{\mathbf{3}} H_{\mathbf{3}} H_{\mathbf{3}} +h.c.\right) +\, \alpha_{2} \, \left(H_{\mathbf{3}} H_{\mathbf{3}} H_{\mathbf{1_1}}+h.c.\right)
   +\, \alpha_{3} \, \left(H_{\mathbf{1_1}} H_{\mathbf{1_1}} H_{\mathbf{1_1}} +\, h.c.\right) ,
\eea
where all the coefficients $\lambda_1, ...,\lambda_{21}$ are non-negative. Besides, $v_{\mathbf{3}}$,\, $v'_{\mathbf{3}}$,\, $v_{\mathbf{1_1}}$,\, $v_{\mathbf{1_3}}$,\, $v''_{\mathbf{3}}$\, and\, $v'_{\mathbf{1_1}}$ are the would-be VEVs for $H_{\mathbf{3}}$,\, $H'_{\mathbf{3}}$,\, $H_{\mathbf{1_1}}$,\, $H_{\mathbf{1_3}}$,\, $\Phi_{\mathbf{3}}$\, and\, $\Phi_{\mathbf{1_1}}$ respectively.
Since the triplets of $T'$ are self-conjugate, any cubic combinations of $\mathbf{3}$ are $T'$ invariant.
The $SU(3)_L$ cubic invariant is either $3 \times 3 \times 3$ or
$\bar{3}\times\bar{3}\times\bar{3}$.  Originally, there are 56 possible Higgs cubic terms.
However, 23 of them violate the $Z_2$ symmetry, 21 of them are either not invariant under $SU(3)_L$ or $U(1)_X$, and 9 of them are not invariant under $T'$. So we are left with only 3 cubic terms. As long as $\alpha_1,\,\alpha_2,\,\alpha_{3}$ are sufficiently less than $\lambda_1, ...,\lambda_{21}$, the cubic terms can be treated as small perturbations and do not alter the pattern of symmetry breaking. The true VEVs should be approximately $v_{\mathbf{3}}$,\, $v'_{\mathbf{3}}$,\, $v_{\mathbf{1_1}}$,\, $v_{\mathbf{1_3}}$,\, $v''_{\mathbf{3}}$\, and\, $v'_{\mathbf{1_1}}$. Since $\lambda_1, ...,\lambda_{21}$ are non-negative, these VEVs minimize the Higgs potential. We plan to return to the issue of perturbations involving these cubic terms in the near future.


\begin{thebibliography}{9}
\bibitem{ATLAS}
  G.~Aad {\it et al.}  [ATLAS Collaboration],
  Phys.\ Lett.\ B {\bf 716}, 1 (2012)
  {\tt arXiv:1207.7214 [hep-ex]}.
\bibitem{CMS}
  S.~Chatrchyan {\it et al.}  [CMS Collaboration],
  Phys.\ Lett.\ B {\bf 716}, 30 (2012)
  {\tt arXiv:1207.7235 [hep-ex]}.
\bibitem{coupling}
ATLAS Collaboration, ATLAS-CONF-2013-034;
CMS Collaboration, CMS PAS HIG-13-005.
\bibitem{CP}
  S.~Chatrchyan {\it et al.}  [CMS Collaboration],
  Phys.\  Rev.\  Lett.\  {\bf 110}, 081803 (2013)
  {\tt arXiv:1212.6639 [hep-ex]}.
  ATLAS Collaboration, ATLAS-CONF-2013-013; CMS Collaboration, CMS PAS
  HIG-13-002.
\bibitem{spin}
ATLAS Collaboration, ATLAS-CONF-2013-031; CMS Collaboration,
CMS PAS HIG-13-003; ATLAS Collaboration, ATLAS-CONF-2013-029
\,and\, ATLAS-CONF-2013-040.
\bibitem{Pontecorvo:1957cp}
  B.~Pontecorvo,
  Sov.\ Phys.\ JETP {\bf 6}, 429 (1957)
  [Zh.\ Eksp.\ Teor.\ Fiz.\  {\bf 33}, 549 (1957)];
  B.~Pontecorvo,
  Sov.\ Phys.\ JETP {\bf 26}, 984 (1968)
  [Zh.\ Eksp.\ Teor.\ Fiz.\  {\bf 53}, 1717 (1967)];
  Z.~Maki, M.~Nakagawa and S.~Sakata,
  Prog.\ Theor.\ Phys.\  {\bf 28}, 870 (1962).
\bibitem{Cabibbo:1963yz}
  N.~Cabibbo,
  Phys.\ Rev.\ Lett.\  {\bf 10}, 531 (1963);
  M.~Kobayashi and T.~Maskawa,
  Prog.\ Theor.\ Phys.\  {\bf 49}, 652 (1973).
\bibitem{Altarelli:2010gt}
  G.~Altarelli and F.~Feruglio,
  Rev.\ Mod.\ Phys.\  {\bf 82}, 2701 (2010)
  {\tt arXiv:1002.0211 [hep-ph]}.
\bibitem{Ishimori:2010au}
  H.~Ishimori, T.~Kobayashi, H.~Ohki, Y.~Shimizu,
H.~Okada and M.~Tanimoto,
  Prog.\ Theor.\ Phys.\ Suppl.\  {\bf 183}, 1 (2010)
  {\tt arXiv:1003.3552 [hep-th]}.
\bibitem{Frampton:1994rk}
  P.~H.~Frampton and T.~W.~Kephart,
  Int.\ J.\ Mod.\ Phys.\ A {\bf 10}, 4689 (1995)
  {\tt hep-ph/9409330}.
\bibitem{Aranda:2000tm}
  A.~Aranda, C.~D.~Carone and R.~F.~Lebed,
  Phys.\ Rev.\ D {\bf 62}, 016009 (2000)
  {\tt hep-ph/0002044}.
\bibitem{Frampton:2007et}
  P.~H.~Frampton and T.~W.~Kephart,
  JHEP {\bf 0709}, 110 (2007)
  {\tt arXiv:0706.1186 [hep-ph]}.
\bibitem{Frampton:2008bz}
  P.~H.~Frampton, T.~W.~Kephart and S.~Matsuzaki,
  Phys.\ Rev.\ D {\bf 78}, 073004 (2008)
  {\tt arXiv:0807.4713 [hep-ph]}.
\bibitem{Eby:2008uc}
  D.~A.~Eby, P.~H.~Frampton and S.~Matsuzaki,
  Phys.\ Lett.\ B {\bf 671}, 386 (2009)
  {\tt arXiv:0810.4899 [hep-ph]}.
\bibitem{Eby:2009ii}
  D.~A.~Eby, P.~H.~Frampton and S.~Matsuzaki,
  Phys.\ Rev.\ D {\bf 80}, 053007 (2009)
  {\tt arXiv:0907.3425 [hep-ph]}.
\bibitem{Ho:2010yp}
  C.~M.~Ho and T.~W.~Kephart,
  Phys.\ Lett.\ B {\bf 687}, 201 (2010)
  {\tt arXiv:1001.3696 [hep-ph]}.
\bibitem{Frampton:2010uw}
  P.~H.~Frampton, C.~M.~Ho, T.~W.~Kephart and S.~Matsuzaki,
  Phys.\ Rev.\ D {\bf 82}, 113007 (2010)
  {\tt arXiv:1009.0307 [hep-ph]}.
\bibitem{Eby:2011ph}
  D.~A.~Eby, P.~H.~Frampton, X.~-G.~He and T.~W.~Kephart,
  Phys.\ Rev.\ D {\bf 84}, 037302 (2011)
  {\tt arXiv:1103.5737 [hep-ph]}
\bibitem{EbyThesis}
 For a recent review of the $\textrm{SM}\times T'$ model see
 D.A. Eby, {\it Binary
Tetrahedral Flavor Symmetry}, Ph D thesis, UNC Chapel Hill, April 2013.
\bibitem{Frampton:1992wt}
  P.~H.~Frampton,
  Phys.\ Rev.\ Lett.\  {\bf 69}, 2889 (1992).
\bibitem{Pisano:1991ee}
  F.~Pisano and V.~Pleitez,
  Phys.\ Rev.\ D {\bf 46}, 410 (1992)
  {\tt hep-ph/9206242}.
\bibitem{Frampton:2012zp}
  P.~H.~Frampton, C.~M.~Ho and T.~W.~Kephart,
  Phys.\ Lett.\ B {\bf 715}, 275 (2012)
  {\tt arXiv:1205.4483 [hep-ph]}.
\bibitem{S3}
  A.~E.~Cárcamo Hern\'{a}ndez, R.~Mart\'{\i}nez and F.~Ochoa,
  {\tt arXiv:1309.6567 [hep-ph]}.
\bibitem{Chen:2007afa}
  M.~-C.~Chen and K.~T.~Mahanthappa,
  Phys.\ Lett.\ B {\bf 652}, 34 (2007)
  {\tt arXiv:0705.0714 [hep-ph]}.
\bibitem{Sen:2007vx}
  S.~Sen,
  Phys.\ Rev.\ D {\bf 76}, 115020 (2007)
  {\tt arXiv:0710.2734 [hep-ph]}.
\bibitem{KraussWilczek}
  L.~M.~Krauss and F.~Wilczek,
  Phys.\ Rev.\ Lett.\  {\bf 62}, 1221 (1989).
\bibitem{IbanezRoss}
  L.~E.~Ibanez and G.~G.~Ross,
  Phys.\ Lett.\ B {\bf 260}, 291 (1991).
\bibitem{LuhnRamond}
  C.~Luhn and P.~Ramond,
  JHEP {\bf 0807}, 085 (2008)
  {\tt arXiv:0805.1736 [hep-ph]}.
\bibitem{Luhn}
  C.~Luhn,
  Phys.\ Lett.\ B {\bf 670}, 390 (2009)
  {\tt arXiv:0807.1749 [hep-ph]}.
\bibitem{EF}
D.A.Eby and P.H. Frampton, Phys. Rev. {\bf D86,} 117304 (2012).
{\tt arXiv:1112.2675[hep-ph]}
\bibitem{Willmann}
L. Willmann, {\it et al.} Phys. Rev. Lett. {\bf 82,} 49 (1999).
\bibitem{Meirose}
B. Meirose and A. A. Nepomucheno,
Phys. Rev. {\bf D84,} 055002 (2011)
{\tt arXiv:1105.6299 [hep-ph]}.
\bibitem{Borges}
J. Sa Borges, Y. A. Coutinho, and E.A. Barreto,
AIP Conf. Proc. {\bf 1520,} 440 (2012).
\end{thebibliography}
\end{document}